\documentclass[12pt,preprint]{aastex}

\slugcomment{}
\shorttitle{Metallicity of the Warm Phase in Early-Type Galaxies}
\shortauthors{Athey et al.}

\begin{document}

\title{Oxygen Metallicity Determinations from Optical Emission Lines in Early-type Galaxies}

\author{Alex E. Athey}
\affil{The University of Texas at Austin}
\affil{Applied Research Laboratories}
\affil{10000 Burnet Rd}
\affil{Austin, TX 78758}
\email{athey@arlut.utexas.edu}

\author{Joel N. Bregman}
\affil{University of Michigan }
\affil{Department of Astronomy}
\affil{500 Church St.}
\affil{Ann Arbor, MI 48109-1090}
\email{jbregman@umich.edu}

\begin{abstract}
We measured the oxygen abundances of the warm (T$\sim 10^{4}K$) phase of gas in seven early-type galaxies through long-slit observations.  A template spectra was constructed from galaxies void of warm gas and subtracted from the emission-line galaxies, allowing for a clean measurement of the nebular lines.  The ratios of the emission lines are consistent with photoionization, which likely originates from the UV flux of post-asymototic giant branch (PAGB) stars.  We employ H II region photoionization models to determine a mean oxygen metallicity of $1.01\pm0.50$ solar for the warm interstellar medium (ISM) in this sample. 
This warm ISM $0.5$ to $1.5$ solar metallicity is consistent with modern determinations of the metallicity in the hot (T$\sim 10^{6}-10^{7}K$) ISM and
the upper range of this warm ISM metallicity is consistent with stellar population metallicity determinations.  
A solar metallicity of the warm ISM favors an internal origin for the warm ISM such as AGB mass loss within the galaxy.

\end{abstract}


\section{Introduction}

\subsection{Warm ISM Discovery}

Early-type galaxies were once thought to contain very little gas in
their interstellar medium (ISM) (e.g. \citet{Galactic_Winds_1,Galactic_Winds_2,Galactic_Winds_3}),
but are now known to host massive amounts of extremely hot (T$\sim 10^{6}-10^{7}K$)
gas that emits in the X-rays \citep{Forman_Einstein}. The reason
early-type galaxies were originally thought to be void of gas is
that  photographic-plate surveys did not reveal the 
warm (T$\sim 10^{4}K$) ISM that is abundant in spiral galaxies
(e.g. \citet{Warm_Gas_Not_Found_2,Warm_ISM_notfound,Warm_Gas_Not_Found_3}).
Only with the advent of large telescopes and efficient detectors was
the warm ISM detected, although the 
masses are an order of magnitude lower than
that of spirals. 
Spectroscopic surveys in the mid-1980's determined warm ISM detection
frequencies in the cores of early-type galaxies of 30\%-50\% \citep{Warm_Spectro_survey_OII,Warm_Spectro_Survey_86_1,Warm_Spectro_Survey_86_2}. 
Subsequent
narrow-band $H\alpha $ imaging studies revealed extended
emission in general agreement with the the stellar morphology and
detection rates consistent with the 50\% rate determined from the early
spectroscopic surveys \citep{Warm_phase_Ha84,Warm_phase_Ha89}.

\subsection{\label{Key-Warm-Phases}Key Warm ISM Studies}
There are three large sample, multi-wavelength studies of early-type 
galaxies that have provided a foundation for our understanding of the cold and warm ISM.
\citet[hereafter TdSA]{Trinchieri_Alighieri} conducted one of the
first quantitative early-type galaxy emission-line studies
with $H\alpha$ imaging and long-slit optical spectroscopy of 13 early-type 
galaxies known to have hot X-ray halos. 
\citep[hereafter G94-5]{Goudfrooij_3,Goudfrooij_2,Goudfrooij_1,Goudfrooij_4}
investigated the origin and evolution of the ISM in an extensive
study of a complete sample of 56 elliptical galaxies using broad-band
imaging, narrow-band $H\alpha $ imaging, long slit spectroscopy,
and \emph{Infrared Astronomical Satellite} (IRAS) far-IR imaging.
A study of the warm and cold phases of the ISM
in early-type galaxies was carried as part of the ESO Key-Program
1-004-43K \citep[hereafter MFCF]{Caon_paper_I,Caon_paper_II,Caon_paper_III,Caon_paper_IV}.
Similar to G94-5, the MFCF study probed the origin
and evolution of the warm and cold ISM phases through a combination of 
broad-band, narrow-band and far-IR
imaging and long-slit spectroscopy of 73 early-type galaxies.  
The two later studies had the benefit of FIR imaging capabilities and added the cold ISM (i.e. dust)
to the discussion of the ISM in early-type galaxies.
These studies sought to determine the frequency of detection of warm and cold ISM in early-type
galaxies and then measure fundamental properties, such as mass, density, and luminosity.  By 
correlating these properties with stellar and hot ISM properties, inferences are made about the
origin and evolution of the ISM and how it relates to the origin and evolution of the galaxy itself.  
Below we briefly summarize the findings of these key studies and note the areas where the studies agree and 
where questions remain.


The three key studies find similar fractions of their samples that host a warm ISM, ranging
from 60\%-80\%, despite the differences in sample selection criteria and sizes of the samples.
TdSA found 75\% of the 13 early-type galaxies surveyed had detectable $H\alpha $
emission which was extended over a range of 5-10 $kpc$. The extent
and elliptical shape of the emission confirmed the galaxy-wide nature of this
phenomena and excluded a nuclear confined process, such as AGN activity.  The TdSA survey was
conducted on galaxies known to have significant masses of hot ISM, and the authors
concede that the high detection rate is likely connected to sample selection. 
The G94-5 optically complete sample (chosen without regard for X-ray luminosity) 
offers a less biased measure of the frequency
of the warm ISM; 60\% of the early-type galaxies surveyed were detected
to have a warm ISM, while a cold (T$\sim 10K$) dust component is
detected in 40\% of these galaxies.  Through the FIR imaging, the G94-5 study was able to show 
that unlike spiral galaxies, early-type
galaxies display a smooth dust component distributed
throughout the galaxy. 
The MFCF authors report an ionized gas detection frequency at
78\% for the 73 galaxies in the optically complete sample.  
The cold dust component was detected 
at a 75\% detection rate with smooth,
extended morphologies similar to that of the warm ISM.  
Two key observations link the warm and the cold ISM together; first, the similar spatial extent 
and morphology of the cold ISM (dust) and warm ISM and second, the fact that the presence
of dust is a strong predictor of the presence of emission lines. It is concluded that these ISM phases
are closely related and likely of the same origin.

The mass of the warm ISM is orders of magnitude below the X-ray emitting hot ISM 
and the stellar mass. 
TdSA calculated a warm ISM gas mass by assuming case-B recombination, a filling factor of $10^{-5}-10^{-6}$,
and densities corresponding to the pressure implied by the X-ray observations.  The resultant 
masses were calculated to be $\sim 10^{5}-10^{6}M_{\sun }$ for
a typical galaxy in their sample.
The $H\alpha$ imaging in G94-5 revealed warm ISM masses similar to the FIR determined dust masses of 
 $\sim 10^{4}-10^{5}M_{\sun }$.
The MFCF study determined masses for the warm ISM in the range of
$10^{3}-10^{5}M_{\sun }$.   Masses of dust at $10^{5}M_{\sun}$ have direct implications on the origin of the
dust as the dust destruction rate is high in the presence of the hot ISM and is further discussed below.

The correlations between the luminosities of $L_{FIR}$, $L_{H\alpha}$,  $L_{X}$, $L_{B}$
and $L_{X}/L_{B}$  explore the links between
the difference phases of the ISM and the stars.  For numerous of reasons, the luminosity correlations have been 
difficult to accurately measure and there are contradictions between the key studies.  
TdSA observed that galaxies
with more massive X-ray halos had stronger line emission,
indicating that the warm gas is somehow linked with
the hot gas.  No other luminosity correlations were observed.
In a related work to G94-5, the authors find a strong anti-correlation between the masses of dust and 
X-ray gas \citep{GThesis,GThesisSummary}. However in a later study with a larger sample (based in part on MFCF data),
 \citet{G97} reveals a positive correlation between the warm ISM with both the hot ISM and the stellar blue luminosity.
MFCF determined the luminosity of the ionized gas 
to be correlated with both blue and X-ray luminosity.  
The scatter in the $L_{H\alpha}$ and $L_{B}$ relation is greatly reduced when 
the blue luminosity is confined to the same region as the emission lines.  The scatter in
all the $L_{H\alpha}$ to $L_{X}/L_{B}$ relations is large and plagued by upper limits on non-detections.

In general, the three studies agree that the ionization source for the warm ISM is
consistent with PAGB stars, but all comment that the data are unable to exclude other excitation sources
such as internal shocks, AGN heating or electron conduction from the hot gas. 
TdSA show that the observed line intensities are
a factor of $\sim 10^{2}-10^{3}$ too large to result from cooling
flow luminosities (e.g. \citet{Cooling_Flows_Fabian91}) and 
conclude that UV flux from post asymptotic giant branch (PAGB) stars are
the most likely ionizing mechanism.
The MFCF observed correlation between $H\alpha$ and blue luminosity is an indication of photoionization
of the warm ISM from PAGB stars. However, the observed correlation with X-ray
luminosity is consistent with warm ISM excitation via electron conduction
from the hot ISM \citep{electron_conduction_Sparks}; 
Neither ionization mechanism is preferred in MFCF's large sample of 73 galaxies.  
See Section \ref{Xmetal_norm_gal} for a continued discussion on the warm ISM ionization source.

The key studies narrow down the origin of this dust and warm gas to either be mass shed from
AGB stars or the result of small galaxy accretion and mergers. 
G94-5 reasons that cold dust 
originated in the circumstellar
envelopes of AGB stars are quickly sputtered
away by the hot ISM with a typical dust grain lifetime ($\sim 10^{7}yrs$) much shorter than the age
of the galaxy \citep{Sputtering}.
And further, the balance between the dust injection and destruction rate 
does not lead to the dust masses inferred by the
far-IR imaging \citep{Faber_Gallagher_Mdot_theory,KGWW}.
The surprising result from the confluence of the G94-5 studies is
that the majority of the dust in elliptical galaxies is external in
origin. Since the distribution of the ionized gas (i.e. the warm ISM) is observed
to be similar to the dust morphology, it is presumed that the warm ISM 
is likewise external to the galaxy, and will
only be seen in galaxies that either have recent merger activity or significant inflows. 
The high detection
frequency of the warm ISM, implies mergers and smaller galaxy acquisitions
are common in early-type galaxies.
Adding to the puzzle, in a subsample of the gas-rich galaxies, the warm ISM was frequently observed
to be kinematically distinct from the main stellar population, also
strongly suggesting an external origin for this gas \citep{Caon_paper_III}.

\subsection{Warm ISM Open Issues}

The warm ISM is important to study in early-type galaxies
and despite the ambitious efforts of the studies described above and subsequent works,
several fundamental questions remain. It is critical to resolve these
open issues since the intertwined relationships between the cold,
warm, and hot gas phases and the stellar population ultimately defines
galaxy evolution. Even though the warm ISM mass is orders
of magnitude less than the stellar mass, discriminating between internal
versus external origin of this material provides pivotal data in
formation models.  Also, even though
minute in mass compared to the cold and hot phases of the ISM, the
optical line emissions of the warm ISM can locally dominate the energy
output for a region. Finally, the optical emission lines of the warm
ISM provides information on the heating and cooling in these systems
which demands further exploration since the observed hot ISM properties
along with the non-detection of cooling flows creates a ``cooling
crisis.''

The metallicity of the warm ISM is unknown. Because the metallicities
of the hot ISM and 
the stellar metallicities (cf. \citet{Trager_1,Trager_2}) are known, determining
the metallicity of the warm ISM can create a link between the gas phases
and the stellar population. Also, the metallicity of the warm
ISM contains important discriminating information concerning the outstanding
issues that the key warm ISM studies were not able to resolve. Specifically,
the line ratios of the ionization species provides information on
the excitation mechanism of the gas; currently, both electron conduction and
photoionization remain viable excitation mechanism. 

Also the metallicity
of the warm ISM contains clues to the internal versus external origin
debate. The gas injected into the ISM from stellar mass loss will generally
be more metal rich than gas accreted from dwarf galaxies (cf. \citet{Z-L_irr_bcg_79}
and \citet{Z-L_dSph_86}). 

Because of the important information contained in the metallicity
of the warm ISM and the unresolved issues concerning its nature and
origin, we have conducted a program to determine the oxygen metallicity
of the $T\sim 10^{4}K$ gas in a small sample of early-type galaxies. Section
\ref{Xmetal Obs Table} describes the sample and observations. 
The data processing techniques and spectral
analysis is discussed in Section \ref{Xmetal_Reductions}. Results
from the emission line analysis is presented in Section \ref{Xmetal Results}.
In Section \ref{Xmetal Discussion}, we conclude with a discussion
of the implications of this work on both the origin of the warm ISM
and the relationships between the stellar population and the other
phases of the ISM.

\section{\label{Xmetal Obs}Observations}

The MDM 2.4 meter Hiltner Telescope coupled with a Boller and Chivens
spectrograph was used to observe 21 nearby, early-type galaxies in four observation
runs over a two year period from 2000 to 2002. The sample was selected to span a range
of hot ISM properties as determined from the \citet{Brown_Bregman} 
X-ray study as well as a range of warm ISM properties determined
in the narrow-band $H\alpha $ imaging surveys by TdSA, G94-5 and
MFCF. The galaxies with their relevant fundamental properties are listed
in Table \ref{Xmetal Gal Prop}. Our warm ISM isolation and detection
method depends on observing galaxies that contain no emission line
gas, therefore,
several galaxies were chosen to have very weak detections of the hot
and warm phases. 

\begin{deluxetable}{ccccccccc}
\tablecaption{Early-type Galaxy Properties: Warm Phase Sample \label{Xmetal Gal Prop}}
\tabletypesize{\scriptsize}
\tablewidth{0pt}
\tablehead {
\colhead{Galaxy} &
\colhead{RA\tablenotemark{a}} &
\colhead{Dec\tablenotemark{a}} &
\colhead{$B_{T}^{0}$\tablenotemark{b}} &
\colhead{D\tablenotemark{b}} &
\colhead{$log\, L_{B}/L_{\odot}$\tablenotemark{b}}&
\colhead{$log\, L_{x}$\tablenotemark{c}}&
\colhead{$log\, L_{H_\alpha }$\tablenotemark{d}}&
\colhead{${H_\alpha }$ Reference}\tablenotemark{e}\\
\colhead{}&
\colhead{(J2000.0)} &
\colhead{(J2000.0)} &
\colhead{}&
\colhead{$(km/s)$}&
\multicolumn{3}{c}{$log ~ (erg ~ s^{-1})$}&
\colhead{}
}
\startdata
NGC 1407&
03 40 11.8&
-18 34 48&
10.57&
$1990\pm 187$&
11.16&
41.34&
39.32&
M96\\
NGC 1600&
04 31 39.8&
-05 05 10&
11.79&
$4019\pm 489$&
10.67&
40.84\tablenotemark{1}&
40.15&
M96\\
NGC 2768&
09 11 37.5&
+60 02 15&
10.93&
$1532\pm 325$&
10.79&
40.41&
\nodata&
\nodata\\
NGC 3115&
10 05 13.9&
-07 43 07&
9.95&
$1021\pm 215$&
10.83&
39.74&
$<37.6$&
M96\\
NGC 3377&
10 47 42.3&
+13 59 08&
11.13&
$857\pm 126$&
10.21&
39.42&
38.95&
G94\\
NGC 3379&
10 47 49.6&
+12 34 55&
10.43&
$857\pm 126$&
10.49&
39.78&
39.04&
M96\\
NGC 3489&
11 00 18.3&
+13 54 05&
11.12\tablenotemark{2}&
$1039\pm101$\tablenotemark{2}&
10.35\tablenotemark{2}&
\nodata&
39.34&
M96\\
NGC 3607&
11 16 54.3&
+18 03 10&
10.53&
$1991\pm 242$&
11.18&
40.82&
39.92&
M96\\
NGC 4125&
12 08 05.8&
+65 10 27&
10.58&
$1986\pm 295$&
11.16&
41.01&
40.30&
T91\\
NGC 4261&
12 42 02.4&
+11 38 48&
11.32&
$2783\pm 590$&
10.35&
41.18\tablenotemark{2}&
39.38&
G94\\
NGC 4374&
12 25 03.7&
+12 53 13&
10.13&
$1333\pm 71$&
10.99&
41.09&
39.56&
G94\\
NGC 4406&
12 26 11.7&
+12 56 46&
9.87&
$1333\pm 71$&
11.10&
41.80&
40.50&
T91\\
NGC 4472&
12 29 46.8&
+08 00 02&
9.32&
$1333\pm 71$&
11.32&
41.77&
39.60&
T91\\
NGC 4494&
12 31 24.1&
+25 46 28&
10.69&
$695\pm 147$&
10.20&
39.28&
\nodata&
\nodata\\
NGC 4552&
12 35 39.8&
+12 33 23&
10.84&
$1333\pm 71$&
10.71&
40.92&
39.26&
M96\\
NGC 4636&
12 42 50.0&
+02 41 17&
10.20&
$1333\pm 71$&
10.96&
41.81&
39.69&
M96\\
NGC 4649&
12 43 39.6&
+11 33 09&
9.77&
$1333\pm 71$&
10.96&
41.48&
39.83&
T91\\
NGC 4697&
12 48 35.9&
-05 48 02&
10.03&
$794\pm 168$&
10.58&
40.13&
39.63&
G94\\
NGC 5044&
13 15 23.9&
-16 23 08&
11.25&
$2982\pm 314$&
10.34&
42.39&
40.73&
M96\\
NGC 5322&
13 49 15.2&
+60 11 26&
11.09&
$1661\pm 352$&
10.80&
40.11&
39.74&
G94\\
NGC 5846&
15 06 29.2&
+01 36 21&
10.67&
$2336\pm 284$&
11.26&
42.01&
40.25&
M96\\
\enddata
\tablenotetext{a}{Values taken from NED (NASA/IPAC Extragalactic Database).}
\tablenotetext{b}{Radial velocity distance corrected for local flows from \citet{Faber_89} unless otherwise noted.}
\tablenotetext{c}{X-ray luminosity from \citet{Brown_Bregman} unless otherwise noted.}
\tablenotetext{d}{H$\alpha$ narrow-band imaging luminosity.}
\tablenotetext{e}{Reference for H$\alpha$ narrow-band imaging luminosity with the following abbreviations: M96 $=$ \citet{Caon_paper_I}, G94 $=$ \citet{Goudfrooij_2}, and T91 $=$  \citet{Trinchieri_Alighieri}. }

\tablenotetext{1}{\citet{Trevor_ROSAT_study}.}
\tablenotetext{2}{RC3, \citet{RC3}.}
\end{deluxetable}

The configuration of the spectrograph was chosen to maximize galaxy
light input with a $2.1\arcsec $ wide by $5\arcmin $ long slit,
while retaining the ability to discriminate between the $H\alpha $
and {[}N II{]}$\lambda 6583$ emission lines. The available and appropriate
grating for these requirements was a 350 lines/mm grating blazed
at $4026$\AA , resulting in $7.1$\AA /pixel and a spectral range
of $\sim 1600$\AA ~ over the 1200x800 pixel Loral CCD. The CCD was characterized
by relatively low read noise at seven electrons with the nominal gain
set at 2.1 electrons per ADU.  The one drawback to the chosen spectrograph 
configuration is that it was necessary to obtain
separate blue and red spectra for each galaxy in order to obtain
all of the important emission lines from {[}O II{]}$\lambda 3727$
to {[}Si II{]}$\lambda \lambda 6717,6731$. The grating is servo controlled
from the control computer and our tests indicated a $5$\AA ~
accuracy in repositioning.  Therefore, internal HgNe and Ne lamps were observed
before and after each grating reposition for wavelength calibration.
Additional calibrations included bias frames, evening and morning
twilight flats when the skies were clear, internal flats illuminated
from an incandescent source, and spectrophotometric standards.

The four observation runs were in March 2000, May 2000, February 2001,
and February 2002. 
Galaxy exposure times and program-type (E $=$ emission-line, T $=$ Template, or L $=$ LINER) are listed in
Table \ref{Xmetal Obs Table}.  The meaning of the program types is discussed below.  During the first three runs, the nights
were cloudy and over half of the full run was lost to weather and instrument problems.  Some
of these observations are of marginal value and only reported here for completeness.  The final run was entirely
photometric.

\begin{deluxetable}{cccc}
\tablecaption{MDM 2.4 Meter Observations\label{Xmetal Obs Table}}
\tabletypesize{\scriptsize}
\tablewidth{0pt}
\tablehead {
\colhead{Galaxy}&
\colhead{Exp Blue (s)}&
\colhead{Exp Red (s)}&
\colhead{Emission Type}\tablenotemark{a}
}
\startdata
NGC 1407&
$8400$&
$6600$&
T\\
NGC 1600&
$8400$&
$5400$&
W\\
NGC 2768&
$12000$&
$11400$&
L\\
NGC 3115&
$6000$&
$4500$&
T\\
NGC 3377&
$7200$&
$6000$&
W\\
NGC 3379&
$6000$&
$6000$&
T\\
NGC 3489&
$6000$&
$4500$&
E\\
NGC 3607&
$9600$&
$7200$&
E\\
NGC 4125&
$12000$&
$10800$&
L\\
NGC 4261&
$6000$&
$4500$&
E\\
NGC 4374&
$6000$&
$4500$&
E\\
NGC 4406&
$9600$&
$6000$&
W\\
NGC 4472&
$9600$&
$6000$&
W\\
NGC 4494&
$1200$&
$0$&
W\\
NGC 4552&
$6000$&
$6600$&
W\\
NGC 4636&
$6000$&
$9900$&
E\\
NGC 4649&
$6000$&
$4500$&
W\\
NGC 4697&
$6000$&
$1800$&
W\\
NGC 5044&
$6000$&
$3600$&
E\\
NGC 5322&
$9600$&
$9900$&
W\\
NGC 5846&
$7200$&
$6600$&
E\\

\enddata
\tablenotetext{a}{Emission Type.  Column display how the observations were ultimately used: T= template galaxy, L=LINER galaxy, E=emission galaxy, W=weak emission.}
\end{deluxetable}

The observing strategy was chosen to suppress spurious emission line
detections and obtain uniform observations of program and template
galaxies. For each galaxy, we attempted to obtain five integrations,
ensuring that reasonable statistics could be employed to eliminate
cosmic rays from the combined data. Because the emission lines are
faint and spread over only a few pixels in both spectral and spatial
directions, a fortuitous single cosmic ray can result in a spurious
warm ISM detection. The total integration time was chosen to result
in a S/N of 5-10 in the template subtracted spectra (See Section \ref{Xmetal_Reductions}). 
We
did not constrain the angle of the slit due to complications
with the instrument rotator during the first two runs; this has little programatic impact
because of the choice
of a wide slit, and the non-preferential orientation of the warm ISM
observed in the narrow-band imaging surveys (TdSA, G94-5, MFCF).

\section{\label{Xmetal_Reductions}Reductions and Spectral Analysis}

The data were reduced in the standard manner using tasks within
IRAF (overscan fitting and subtraction, bias frame construction and subtraction).
Each of the runs was calibrated separately but in a similar manner.
Internal lamps were used to create a response image and divided
through the data, eliminating differences in pixel-to-pixel sensitivity
on the CCD.  Twilight flats were used to construct an illumination image,
correcting for large scale structure and slit illumination. For the first two runs, the incandescent internal lamp produced
too few photons in the far blue and only added noise to the data.
For these runs we used wavelength calibrated twilight flats to
produce response images. A HgNe lamp was used for wavelength calibration of blue-tuned observations
and a Ne lamp was used for red-tuned observations. Residuals in the linear,
spectral solution were much better (typically $<1$\AA ) than the
resolution element ($\sim 8$\AA ) over the entire chip.

The galaxies were summed over the spatial dimension based on the visual
inspection of the 2-D spectra, with background regions selected from
the outer regions of the chip. Typically the summed
regions for the galaxies were about $1.2\arcmin $ of the $5.2\arcmin $
unvignetted slit length. The spectra were flux calibrated from observations
of spectrophotometric standards \citep{Spectrophoto_stds_Massey88}
and corrected for extinction with the standard correction for Kitt
Peak distributed with IRAF 2.11.

\begin{figure}
\epsscale{0.75}
\includegraphics[angle=-90,scale=.65]{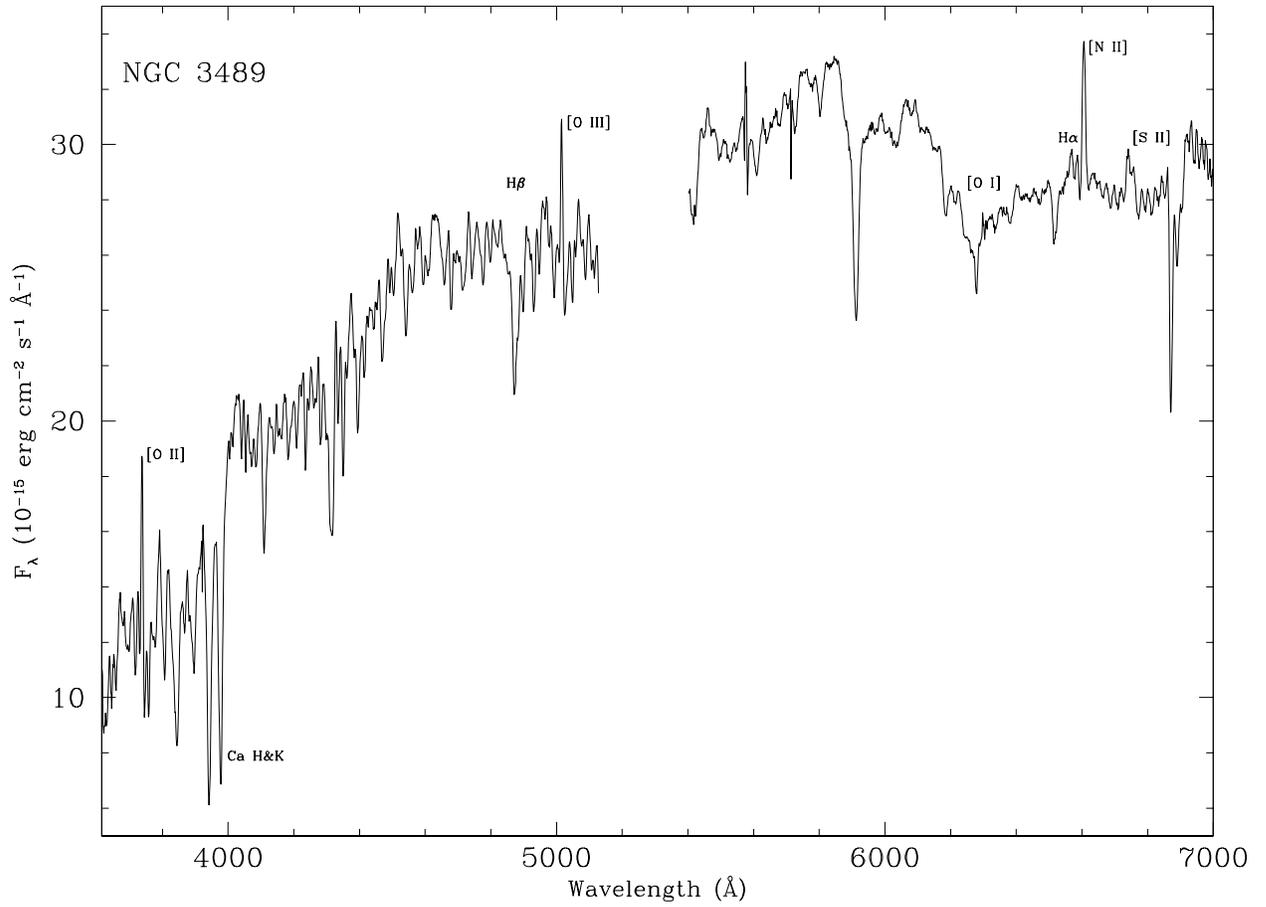}
\caption[Sample Emission-Line Galaxy Optical Spectra: NGC3489]
{Composite Blue and Red Spectra of NGC3489.  Important emission lines
that are visible in the un-subtracted data are marked aobve the continuum
level.  In addition, calcium H \& K lines are marked for reference.}
\label{xmetal_raw_spec}
\end{figure}

Custom code was developed to implement the combining of the individual
1-D spectra, template matching and subtraction, as well as the extraction
of line fluxes. The 1-D spectra combination procedure employed an iterative
$1.5\sigma $ rejection algorithm which eliminated cosmic rays that
were overlooked in the aperture summing rejection methods. The wavelength
calibration was tuned to night sky lines and the measured
redshift was checked and tuned to strong stellar population lines
(e.g. Ca H\&K in the blue and {[}N II{]}$\lambda 6583$ in galaxies
showing emission lines in the red). An example summed spectra is shown
in Figure \ref{xmetal_raw_spec}.

Although we have upper-limits from the $H\alpha $ imaging surveys,
it is unknown which galaxies are void of warm ISM emission. Therefore, the
determination of template galaxies, or galaxies without any detected
warm ISM, is an involved and iterative process. First the $H\alpha $ and {[}N II{]}$\lambda 6583$
region of the spectra is examined for obvious emission. Using the
local continuum, upper limits on the line fluxes are determined for
galaxies with no obvious $H\alpha $ and {[}N II{]}$\lambda 6583$
emission.  
Ten of
the original 21 galaxies match this criteria and act as potential
template galaxies. We take these ten galaxies and combine
them into a master template galaxy. When combined into the template galaxy, the strongest
of these weak emitters will not be included in the template creation because of the aggressive clipping
algorithm in the combination routine. 
To determine the the weak emitters 
the template is subtracted from the individual galaxy observations
and examined for emission lines.
Three of the ten template galaxies revealed oxygen lines and were removed from
the master template. This process was repeated with the emitters removed and reveal 
four more weak emitters, leaving three galaxies without emission line activity: NGC 1407,
NGC 3115, NGC 3379.  As
an independent check of our template definition methodology, 
all seven of the potential template galaxies were subtracted from the galaxy with
the strongest emission lines (NGC 2768) and we determined that the three template
galaxies selected resulted in the high reported fluxes (i.e. lowest
background). 

The three template galaxies, 
NGC 1407, NGC 3115 and NGC 3379, 
lack of warm emission is consistent with results from the narrow-band
imaging detections for these galaxies. NGC 3379 is reported to have
a weak $H\alpha $ in MFCF but an upper limit below this reported
detection is determined in G94-5. NGC 3115 is detected only as an
upper limit by MFCF and has been previously used as a stellar template
galaxy by the Palomar spectral survey of the nuclear regions of 500
nearby galaxies \citep{Luis_LINER_ngc3115_template}. NGC 1407 is
marginally detected in both the G94-5 and MFCF studies, but the fluxes
reported differ by a factor of five. The inconsistencies of weak detections
in the narrow-band surveys reveal intrinsic limitations of the narrow-band
technique. Broad-band red imaging is used to subtract an
underlying stellar population from the narrow-band data; a process which is
susceptible to scaling errors.  The stellar template constructed from our sample
contains $<10^{-16} ~ erg\, s^{-1}\, cm^{-2}$ combined $H\alpha $
and {[}N II{]}$\lambda 6583$ flux; this is an order of magnitude more sensitive
as a non-detection than the narrow-band imaging surveys. 

\begin{figure}
\epsscale{0.75}
\includegraphics[angle=-90,scale=.31]{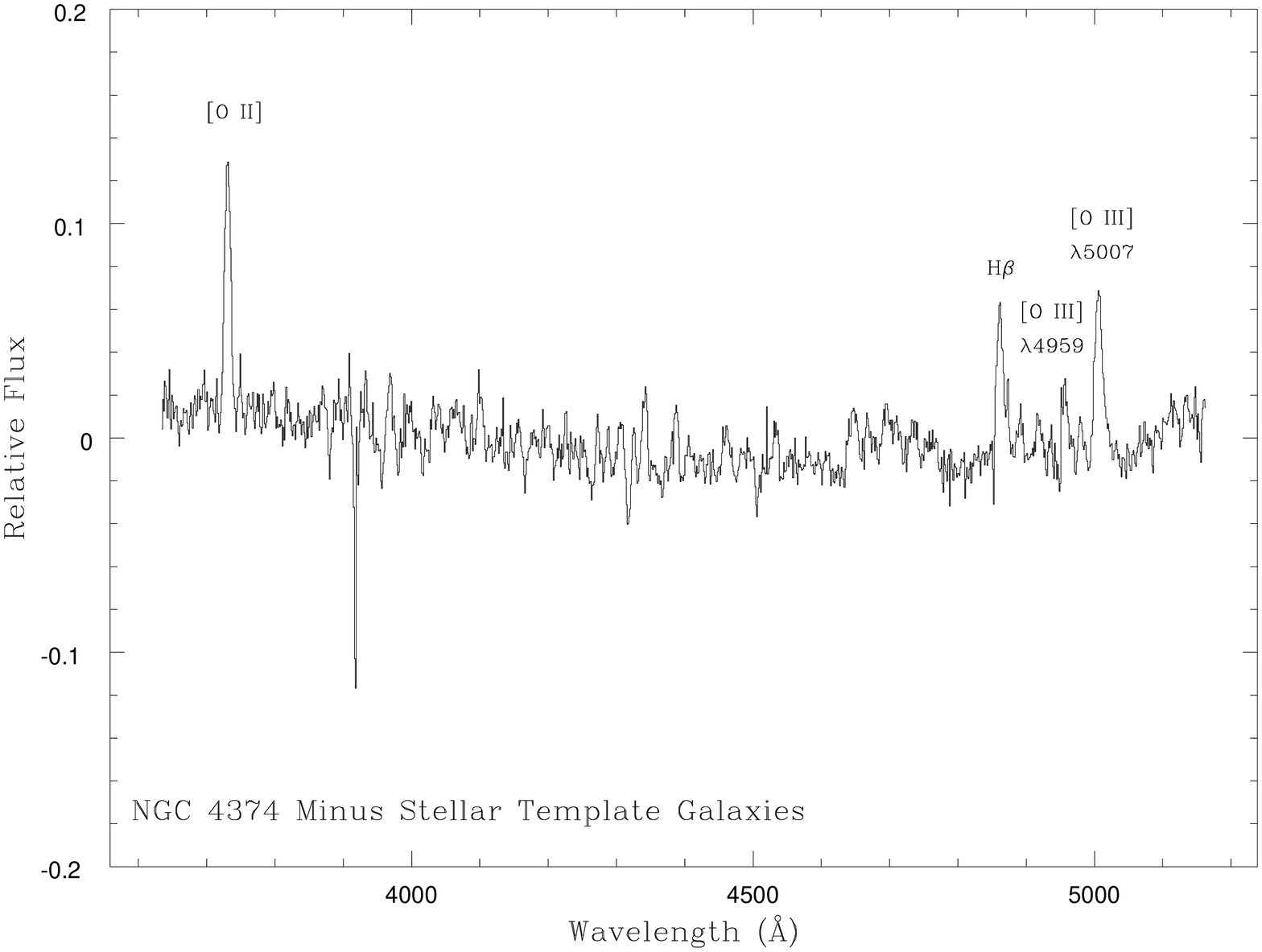}
\includegraphics[angle=-90,scale=.31]{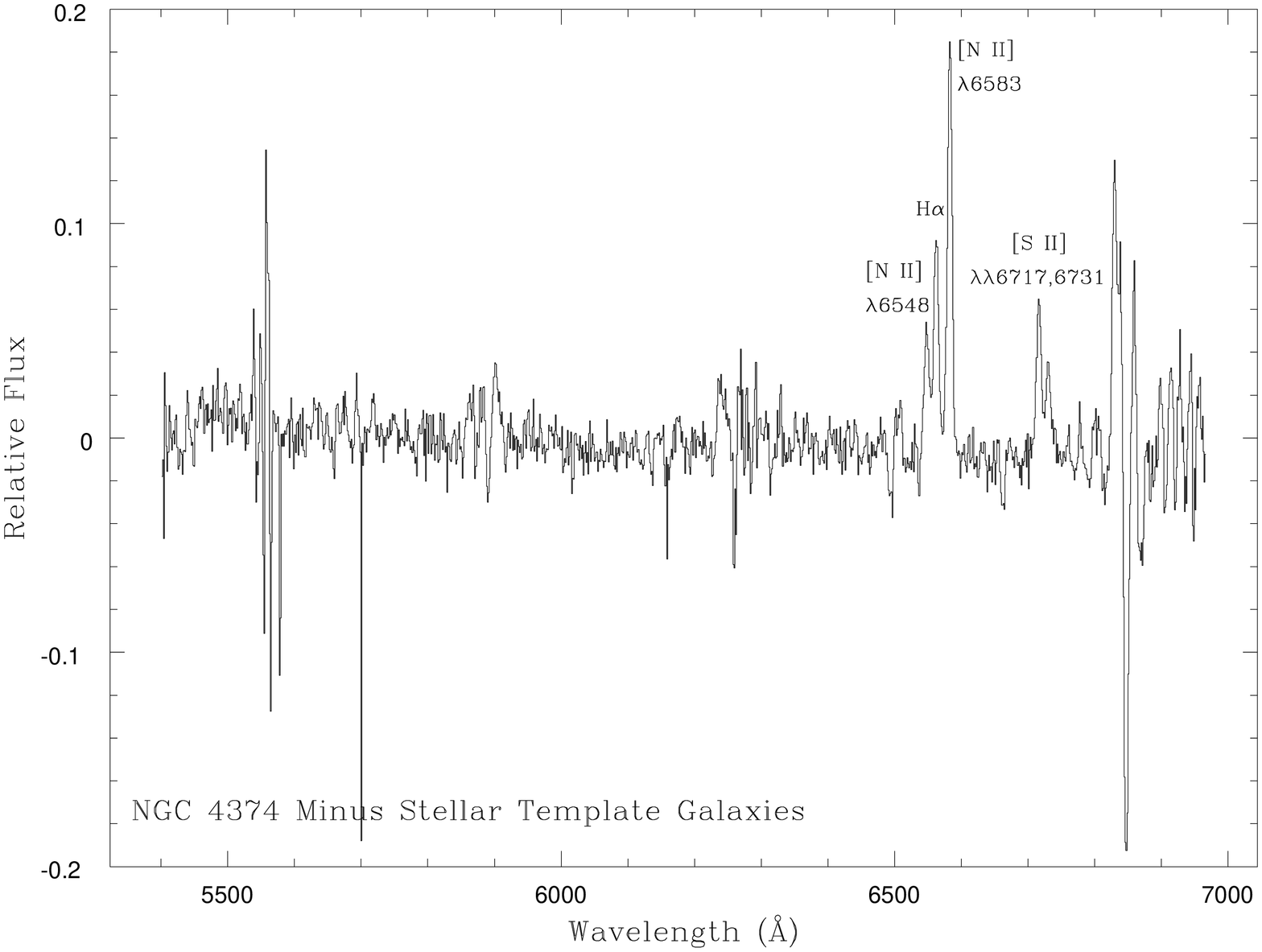}
\caption[Sample Emission-Line Galaxy Subtraction - Blue: NGC4374]
{Blue (left) and Red (right) spectra of NGC4374 with template galaxies subtracted, removing the stellar
continuum. Large residuals at 5500\AA \ and 6800\AA \  are due different night sky lines positions in the rest wavelength
of NGC4374 and template galaxies. The large residual at 3900\AA \ is due to the imperfect matching between the Ca H\&K lines of the template galaxies and NGC 4374.}
\label{xmetal_sub_blue}
\end{figure}

Once the template galaxies are defined, the stellar population was
subtracted from each of the other galaxies, revealing emission
lines and any differences in stellar populations. An example subtraction
for NGC 4374 is shown in Figure \ref{xmetal_sub_blue}. 
In all galaxies, the residuals
in the subtractions are flat with large-scale variation at a 8-10\% level in the blue and better
than 5\% in the red. Once the subtractions are made, the relative line
fluxes are measured by defining a low order polynomial to describe
the local background and using a gaussian profile to extract the line
flux. A gaussian decomposition is only necessary in the crowded
{[}N II{]}$\lambda 6548$, $H\alpha $, {[}N II{]}$\lambda 6583$
region and the fitting method gives equivalent results to raw counts above
a background for all other lines. Note that because of the coarse
spectral sampling and the relatively low velocity dispersions,
the width of the gaussian profiles contain no useful kinematic information
and is thus not reported. 
In eight of the non-template galaxies, the signal-to-noise 
in the data are insufficient for
emission line work, or only H$\alpha$ is observed.
For the rest of the sample, 
extracted relative line
fluxes normalized to $H\beta $ are reported in Table \ref{Xmetal Line Ratios}.
The reported flux ratios have been dereddened based on an
average value for the reddening from \citet{Reddening_Burnstein_Heiles}
and \citet{Reddening_Schlegel}. The observed relative fluxes reported here do not correct for
internal galaxy reddening, displaying the Balmer decrements as observed; Below
we apply the correction (See Section \ref{Xmetal WP Z}). 
The errors reported are determined
by calculating the range of acceptable background levels within the
noise of the local continuum for isolated lines and combining this
background uncertainty in quadrature with the range of acceptable
gaussian widths that reproduce the total observed line flux for the
crowded {[}N II{]}-$H\alpha $ complex.

\begin{deluxetable}{ccccccccccccc}
\tablecaption{Observed Nebular Emission Line Ratios Relative to H$\beta$ \label{Xmetal Line Ratios}}
\tabletypesize{\scriptsize}
\tablewidth{0pt}
\tablehead {
\colhead{Galaxy}&
\colhead{[O II]}&
\colhead{[Ne III]}&
\colhead{$H\beta $}&
\colhead{[O III]}&
\colhead{[O III]}&
\colhead{[O I]}&
\colhead{[N II]}&
\colhead{$H\alpha $}&
\colhead{[N II]}&
\colhead{[S II]}&
\\
&
\colhead{$\lambda 3727$}&
\colhead{$\lambda 3869$}&
\colhead{$\lambda 4862$}&
\colhead{$\lambda 4959$}&
\colhead{$\lambda 5007$}&
\colhead{$\lambda 6300$}&
\colhead{$\lambda 6548$}&
\colhead{$\lambda 6563$}&
\colhead{$\lambda 6583$}&
\colhead{$\lambda \lambda 6717,6731$}
}
\startdata
NGC 2768&
$70\pm 7$&
$14\pm 6$&
$10\pm 3$&
$1\pm 4$&
$24\pm 4$&
$8\pm 2$&
$20\pm 4$&
$29\pm 5$&
$45\pm 7$&
$50\pm 8$\\
NGC 3489&
$19\pm 5$&
$2\pm 3$&
$10\pm 2$&
$2\pm 5$&
$13\pm 5$&
$3\pm 2$&
$30\pm 8$&
$34\pm 5$&
$57\pm 8$&
$37\pm 7$\\
NGC 3607&
$32\pm 8$&
$6\pm 3$&
$10\pm 3$&
$7\pm 3$&
$20\pm 4$&
$2\pm 4$&
$25\pm 5$&
$48\pm 6$&
$59\pm 8$&
$22\pm 5$\\
NGC 4125&
$74\pm 8$&
$14\pm 4$&
$10\pm 2$&
$23\pm 5$&
$71\pm 10$&
$3\pm 3$&
$19\pm 4$&
$34\pm 8$&
$40\pm 5$&
$23\pm 7$\\
NGC 4261&
$16\pm 4$&
$6\pm 5$&
$10\pm 3$&
$6\pm 2$&
$12\pm 3$&
$3\pm 2$&
$15\pm 5$&
$22\pm 8$&
$39\pm 5$&
$15\pm 6$\\
NGC 4374&
$16\pm 4$&
$<1$&
$10\pm 2$&
$5\pm 2$&
$12\pm 4$&
$2\pm 4$&
$21\pm 9$&
$35\pm 6$&
$49\pm 8$&
$29\pm 5$\\
NGC 4636&
$24\pm 5$&
$2\pm 4$&
$10\pm 3$&
$4\pm 4$&
$13\pm 3$&
$2\pm 5$&
$10\pm 3$&
$33\pm 5$&
$49\pm 7$&
$21\pm 5$\\
NGC 5044&
$24\pm 3$&
$2\pm 3$&
$10\pm 2$&
$6\pm 2$&
$20\pm 3$&
$3\pm 2$&
$33\pm 8$&
$32\pm 6$&
$53\pm 8$&
$33\pm 5$\\
NGC 5846&
$26\pm 5$&
$2\pm 3$&
$10\pm 4$&
$6\pm 3$&
$14\pm 4$&
$<1$&
$8\pm 5$&
$29\pm 4$&
$36\pm 8$&
$13\pm 10$\\
\enddata
\end{deluxetable}

\section{\label{Xmetal Results}Warm ISM Emission Line Results}

\subsection{\label{Xmetal Excitation Mech}Excitation Mechanism}

\subsubsection{\label{Xmetal_norm_gal}Normal Galaxies}

There are many possible ionization mechanisms for the warm ISM that
have been previously explored and ruled out by the key warm ISM studies 
(Section \ref{Key-Warm-Phases}) and other works.  Photoionization by hot young
stars \citep{Warm_phase_Ha89,Warm_Phase_YSO_Ionization} is inconsistent
with both the optical colors (G94-5; MFCF) and integrated optical
spectra \citep{WP_Sadler87,WP_Heckman88}. Ionization from radiation
associated with an AGN \citep{Warm_Phase_AGN_Ionization} is ruled
out by the large radial extent of the warm ISM observed in the narrow-band
imaging (TdSA; G94-5; MFCF). Shock excitation \citep{Warm_Phase_Shock_Ionizaion_79,Warm_Phase_Shocks_Ionization}
would produce filamentary ionization regions, where smooth
emission is observed (TdSA; G94-5; MFCF). Further, the the input energy
from shocks is too low by two orders of magnitude for typical densities
and velocities as noted by \citet{electron_conduction_Sparks}. Condensation
out of the hot phase and into warm phase filaments has been investigated
for a number of X-ray clusters \citep{Warm_Phase_Condensation_Clusters_80,Warm_Phase_Condensation_Clusters_85}
but the large number of recombinations required per hydrogen atom
(e.g. \citet{WP_Johnstone87}; TdSA and references therein) is inconsistent
with the observed recombination rates for early-type galaxies \citep{electron_conduction_Sparks}.

The three remaining viable excitation mechanisms for the warm ISM
are photoionization from PAGB stars, electron conduction
from the hot phase (e.g. \citet{electron_conduction_Sparks}), and
photoionization by extreme UV (EUV) photons from the hot phase \citep{Donahue_Voit_90,Donahue_Voit_91}.
The data in the key warm ISM studies are equally consistent with all
of these ionization mechanisms.  However, photoionization
by PAGB stars is the preferred mechanism since these old stars are known
to populate the galaxies and given the  significant correlation
of $L_{H\alpha}$ with $L_{B}$ within a given region (MFCF) strongly indicates a stellar ionizing source.  Further, the
 presence of the far ultraviolet
(UV) flux has been confirmed in a small sample of 30 early-type galaxies
with \emph{International Ultraviolet Explorer} (IUE) 
(\citet{UV_ETGs_IUE_Burnstein88};  For a review of Far UV flux in early-type
galaxies see \citet{UV_ETG_Review_OConnell99}).  

However, to adopt PAGB stars as the ionizing source, excludes the observed correlations
between $H\alpha $ and X-ray luminosity (TdSA; \citep{G97}, MFCF).
The correlations between the X-ray and warm ISM morphologies \citep{WP_X_morph_link,WP_HP_link_TNdSA_97,WP_HP_link_TG_02}
provide additional support for photoionization of the warm ISM by the
EUV flux from the hot ISM, but it is possible to construct 
physical conditions that reproduce these morphological links (e.g. cooling
of the hot ISM).  Although there have been detected correlations between the 
$L_{H\alpha}$ and $L_{X}/L_{B}$, the correlations contain much more scatter and
are less significant than the correlation between $L_{H\alpha}$ with $L_{B}$ within a given region.  The
large amount of scatter in the X-ray relationship and the known high $L_{X}/L_{B}$ galaxies
which contain weak or no $H\alpha$ emission and strong $H\alpha$ emitters with low  $L_{X}/L_{B}$ 
provide insight that the hot and warm ISM link is more tenuous than the link between the 
stars and the warm ISM.  Finally, the key studies in in Section \ref{Key-Warm-Phases} and subsequent
work \citep{SauronV}, indicate a strong connection between the dust (cold ISM) and 
the emission line regions, but there are not strong correlations between $L_{FIR}$ and $L_{X}/L_{B}$ 
in any of the key studies or in recent Spitzer Observations \citep{Temi2007}.

Emission line ratios of the different ionization species have the
ability to discriminate between collisionally excitation and photoionization.
Comparing our sample's line ratios to line ratios predicted for photoionization
models of \citet{Photo_1}, \citet{Photo_2}, and \citet{Photo_3}
and shock ionization of \citet{AllenGroves2008}, we conclude the data
are consistent with photoionization.  Unfortunately, these data are not discriminating due to
the large errors in the data and the region populated by these galaxies happens to land in an overlap region
for collisional and photoionization.



Additional support for the photoionization of this gas is seen with
the Far Ultraviolet Space Explorer (FUSE) \citep{Joel_FUSE}. The observed UV radiation field
produces sufficient photons shortward of 912\AA \ to ionize the warm ISM and is consistent
with the number of predicted extreme horizontal-branch stars, or equivalently
PAGB stars. 

When the data presented here is combined with the literature, it presents a
reasonable case for photoionization of the warm ISM.  This is the preferred 
ionization mechanism in the literature and in discriminating
between the two sources of photoionization, PAGB stars or the hot
ISM, it would seem more probable that PAGB stars provide the necessary ionizing 
photons.

\subsubsection{LINER Galaxies}

Four of the galaxies listed in Table \ref{Xmetal Line Ratios}
are reported as low ionization nuclear emission-line regions (LINERs)
galaxies in the literature: NGC 2768 \citep{LINER_Heckman80}, NGC
3115 \citep{Luis_LINER_ngc3115_template}, and NGC 4125
and NGC 4261 \citep{LINER_SPEC_NGC4261}. The question of identical
treatment of these LINER galaxies and the normal galaxies arises.
LINERs are strictly defined to be galaxies with line ratios of {[}O II{]}$\lambda 3727$
greater than {[}O III{]}$\lambda 5007$ and {[}O I{]}$\lambda 6100$ greater than
one-third {[}O III{]}$\lambda 5007$ \citep{LINER_Heckman80}. While
useful for separating classes of objects, the LINER definition as
noted by its creator, Heckman, has no physical basis and in reality only provides
a crude boundary between photoionization and a low-level
AGN activity.

For our purposes, the LINER identification acts as a flag for further investigation,
and not as an automatic exclusion filter.
The line ratios for the four LINER galaxies are consistent within the photoionization, but the excitation
mechanism and thus the valid regions in diagnostic diagrams are
still uncertain and being debated for LINERs (for a review of the
physical mechanisms behind LINERs see \citet{LINER_review_96} and
\citet{LINER_review_02}). Although it is not entirely clear what
to expect for LINER galaxies, the presence of strong collisionally
excited lines casts a cloud of doubt of pure photoionization. 
The {[}Ne III{]}$\lambda 3869$ line
is an indicator of the importance of collisional excitation because
the collisional cross-section is over an order of magnitude
larger than the photoionization transition probability for this line. NGC 2768
and NGC 4125 have significant {[}Ne III{]}$\lambda 3869$ fluxes and
thus we conclude that these systems have a more complex physical structure
and excitation mechanism than photoionization. These two galaxies
are excluded in the subsequent sections where pure photoionization is assumed.
Note that NGC 4125 and NGC 4261 show no distinguishing line ratios
from the normal galaxies and, the data presented here do not
support the LINER classification.  These two galaxies were classified as LINERs 
by \citep{LINER_SPEC_NGC4261} which had a similar observational aperture as this study, but
only had wavelength coverage down to 4200\AA, excluding the useful {[}O II{]}$\lambda 3727$ information.

\subsection{\label{Xmetal WP Z}Oxygen Metallicity}

For those galaxies where photoionization is the dominant ionization mechanism, the
line ratios can be used to determine the metallicity of the warm phase
gas. There are two critical assumptions to be addressed before using these line ratios and photoionization
models to determine global metallicities.  The first assumption is that 
UV flux from PAGB stars ionizing a diffuse ISM on a galactic scale is well represented
by photoionization models that were developed for Str$\ddot{{o}}$mgren-sphere-type
H II regions.  First, we note that the models are determining  
the balance between photoionization by UV photons and recombinations to the ions; this
is fundamentally the same equation in a galaxy-wide ISM as an HII region, given a source
of ionizing photons internal to a volume of gas.  The UV flux and electron densities of the two environments
are similar, although, PAGB stars have different spectral shapes than
high mass main sequence stars, but \citet{R23_Stacy91} shows that the photoionization models are 
largely insensitive to the spectral shape of the ionizing star.    
The models assume the gas is ejected from the ionizing star (and thus
of the same metallicity), this is a feature used to calibrate the observed line ratios, 
but a direct link between the ionizing source and the ejecta is not a necessary 
condition to employ the models.  We argue that the photoionization models will be applicable to the
warm ISM, but we readily acknowledge that there are differences in these two environments which will
lead to will lead to greater uncertainties in its application here. 

The second key assumption is that the analysis of the linear
combination of many individual diffuse ionized regions observed in an integrated
galaxy spectra produces a result that is indicative of the mean properties of the observed 
regions.  \citet{Global_HII_Diagnostics}
conducted a study addressing these issues in nearby galaxies, observing
H II regions individually through small apertures and globally with
drift scanning with a long slit. The authors conclude that global
metallicity determinations, accurate to $0.2$ dex, are possible with long-slit, integrated spectra.

\begin{deluxetable}{cccc}
\tablecaption{R23 parameter and Oxygen Metallicity\label{Xmetal Z}}
\tabletypesize{\scriptsize}
\tablewidth{0pt}
\tablehead {
\colhead{Galaxy}&
\colhead{log R23}&
\colhead{[O/H]}&
\colhead{$Z_{Oxygen}/Z_{\sun}$}
}
\startdata
NGC 3489&
$0.63^{+0.11}_{-0.14}$&
$-3.26 \pm 0.2$&
$1.13^{+0.49}_{-0.30}$\\
NGC 3607&
$0.93^{+0.08}_{-0.11}$&
$-3.66 \pm 0.2$&
$0.43^{+0.29}_{-0.18}$\\
NGC 4261&
$0.51^{+0.06}_{-0.07}$&
$-3.14 \pm 0.2$&
$1.44^{+0.47}_{-0.29}$\\
NGC 4374&
$0.58^{+0.09}_{-0.10}$&
$-3.20 \pm 0.2$&
$1.26^{+0.50}_{-0.31}$\\
NGC 4636&
$0.68^{+0.09}_{-0.11}$&
$-3.31 \pm 0.2$&
$0.98^{+0.41}_{-0.26}$\\
NGC 5044&
$0.75^{+0.05}_{-0.06}$&
$-3.37 \pm 0.2$&
$0.86^{+0.36}_{-0.24}$\\
NGC 5846&
$0.68^{+0.08}_{-0.09}$&
$-3.30 \pm 0.2 $&
$1.00^{+0.37}_{-0.26}$\\
\enddata
\end{deluxetable}

Metallicity determinations of ionized gas in individual H II regions
are built upon photoionization
models, determining ratios of ionized species relative to hydrogen.
The most common diagnostic is the $R23$ parameterization, $R23\equiv (([O\, II]\, \lambda 3727+[O\, III]\, \lambda \lambda 4959,\, 5007)/H\beta) $, 
defined by \citet{R23_Pagel79} and refined by \citet{R23_Stacy91}.
This parameterization is convenient since it employs the readily available
optical oxygen lines, however,  it requires the {[}O II{]}$\lambda 3727$/{[}N
II{]}$\lambda 6583$ information to resolve a degeneracy present in the $R23$ models.

We take the line ratios reported in Table \ref{Xmetal Line Ratios} and correct for reddening
internal to the galaxy using the interstellar extinction curve of \citet{Savage79} and
assuming case B ratio for the $H\alpha$ to $H\beta$ at $T=10^{4}$ \citep{Osterbrock89}.
For NGC 4261 no reddening correction is applied, as the Balmer decrement indicates
a non-physical negative reddening.

These dereddened lines are used to 
generate $R23$ values for the galaxies.
Because of the faint nature of the $[O\, III]\, \lambda 4959$ line,
we used the conical, equilibrium value of 1/3 the line flux of $[O\, III]\, \lambda 5007$
when this line was buried in the noise of the subtracted spectra.
All of the galaxies end up on the upper branch of the $R23$ curve as indicated by
the  {[}O II{]}$\lambda 3727$/{[}NII{]}$\lambda 6583$ diagnostic.  
To determine metallicities, we
employ \citet{R23_Stacy91} calibration as parameterized by \citet{Global_HII_Diagnostics} and
reported the metallicity results in Table
\ref{Xmetal Z}.
For the zero point of the solar metallicity
we employ a solar photospheric oxygen abundance of {[}O/H{]}$=-3.3$ 
\citep{Abundances_new_Oxygen,Abundances_newest_Oxygen}, but note
the uncertainty and current debate about the solar determinations. 
The errors reported on the oxygen metallicities in these galaxies
are at least $0.2$ dex as these are the dominating and intrinsic uncertainties
in the models used to create the $R23$ grid and its application to
global galaxy spectra (cf. \citet{Global_HII_Diagnostics}). The
median metallicity for the seven early-type galaxies
is O/O$_{\sun }$=1.01. Within
these $0.2$ dex absolute errors, the none of the galaxies are significantly
different from one another.

\section{\label{Xmetal Discussion}Implications and Discussion}

The determined mean oxygen metallicity of solar for the warm phase of the ISM has
implications for the origin of this gas and its relationship to
other phases of the ISM and the stellar population.  The limitations of this study
are apparent with a small sample of seven
galaxies, large $0.2$ dex errors, and a less than ideal application of H II region models to the warm
ISM.  However, this work does present new information as it appears that there
are no determination of the metallicity of the warm ISM in early-type galaxies in the literature.
\citet{SauronV} conducted the most recent large survey of ionized gas in early-type
galaxies and conclude at the end of the study that the origin of the ionized gas is still 
unsolved and state that the metallicity provides an insightful clue.  Below we
consider the implications of a near solar metallicity for the warm ISM.

\subsection{\label{Xmetal WP Origin}Origin of the Warm ISM}

A warm ISM with a solar metallicity argues against
an external origin model for the emission line gas. The accretion galaxies involved in 
minor mergers are presumably dwarf irregular (dIrr)
galaxies, since dwarf spheroidal (dSph) galaxies are most frequently
devoid of gas.  Gaseous and dusty galaxies with near solar metallicity are luminous 
($>10^{10}L_{\sun}$) and massive\citep{L-Z_dIrr_dSph}.  Mergers of these types of objects
produce profound disturbances, which are simply not observed in these early-type galaxies with $H\alpha$ emission. 
It is possible that an intense gas enrichment by supernovae type-II
occurs once low metallicity gas is accreted,
however, the broad-band
colors and global galaxy spectra (G94-5, MFCF) show little evidence
for major star formation.  Further, the observed supernovae rates are too low to
account for such a drastic enrichment \citep{Cappellaro_SN_rates}.

We can compare a solar metallicity for the warm ISM with metallicities of the hot ISM and the stars.
The stellar population in early-type galaxies is generally observed to have a
solar or super-solar metallicity \citep{Trager_1,Trager_2}.  
A number of the seven galaxies observed in this work have stellar Fe and $\alpha$ metallicity 
measurements in the literature, although, there are some inconsistencies between
authors (NGC 4261 is measured to have $[Fe/H]$ of -0.03, 0.275, 0.29, and 0.188 by
\citet{Trager_1}, \citet{Thomas2005}, \citet{Howell2005}, and \citet{SanchezBlazquez2006},
respectively.)  Not all the galaxies have measured stellar metallicities but a five galaxy average 
from all of the determinations in the 2005 and 2006 papers listed above is $[Fe/H]=0.22$, ranging
from $[Fe/H]=0.099$ to $[Fe/H]=0.44$ ($[\alpha/H]$ is similar).  
It is important to remember that \citet{Trager_2} estimates the \emph{absolute} error on 
stellar metallicity determinations close to a factor of two, while the relative errors are much smaller.  
The \emph{Chandra} X-ray Observatory has the spatial resolution to resolve and remove the stellar component from the gas which is
leading to better data and a revision of our (mis)understanding of the metallicity in the hot ISM.  Recently,
\citet{HumphreyBuote2006} analyzed a sample of 28 early-type galaxies observed with \emph{Chandra} and determine that 
the hot ISM abundances are consistent and perhaps higher than the stellar abundances.  
In summary, a $0.50$ to $1.50$ solar metallicity in the warm ISM  is consistent with both hot ISM and stellar metallicity determinations.  

It has been thought that early-type galaxies would have a problem generating and retaining dust
and warm gas, due to the hostile conditions provided by the hot ISM and the low injection rate
of new material into the ISM from stars.
\citet{Parriott2008} have shown that is it possible for dust to condense in 
the circumstellar envelopes of AGB stars and
then subsequently be injected into the hot ISM. 
Although this dust has a short lifetime in the hot ISM, the injection rate 
has been observed to be greater than the destruction
rate and could account for the total observed mass given a simple evolution
of this injection rate given that more massive stars evolved off the main sequence in the past  \citep{ISO_Paper}.
An internal origin for the dust is supported by
the G94-5 and MFCF studies indicating that the majority ($\sim 90\%$) of
the dust is distributed smoothly throughout the galaxy. The internal
origin for the cold ISM is further confirmed by a 2.29$\mu $m
CO absorption feature study that finds a lower velocity dispersion
for the dust than that of the stars in a sample of 25 nearby early-type
galaxies \citep{CP_Dust_velocity_dispersion}.  Finally from the analysis of
\emph{Spitzer} observations of 46 galaxies in the FIR ($24\mu$m, $70\mu$m, and
$160\mu$m), \citet{Temi2007} 
propose an AGB mass loss, internal origin for the observed dust,
which sinks to the core of the galaxy and is periodically driven out by AGN activity.  This 
outward transport of dust may cool the hot gas down to the warm phase.  

The major stumbling block to a galaxy internal origin for dust and gas is the observations of
gas kinematics which are distinct with respect to the stars.  Recent observational evidence has suggested that
there may be different evolutionary tracks for early-type galaxies. The classification of early-type
galaxies into fast and slow rotation categories by \citet{Sauron9} and the key ingredient of gas
for the formation and evolution of fast rotators provides an important dividing line for future discussions 
on the origin of the ISM in early-type galaxies.

\acknowledgements{
The authors would like to acknowledge useful discussions with P. Goudfrooij and E. D. Miller.  We would also like to thank an 
anonymous referee for suggestions that significantly improved this paper.
\bibliographystyle{apj}
\bibliography{thesis}

\end{document}